**High-Efficiency and Large Supercurrent Rectification with Superconducting Coil**

*Takumi Ichikawa, Daisuke Takegami*, Gen Nishijima, Shunsuke Kawakami, Tsutomu Nojima*, Yuji Tsuchiya, Satoshi Awaji, Fuyuki Ando, Ken-ichi Uchida, Yuya Hattori, Yoshikazu Mizuguchi**

Takumi Ichikawa, Daisuke Takegami, Yuya Hattori, Yoshikazu Mizuguchi
Department of Physics, Tokyo Metropolitan University, Hachioji, Japan
E-mail: dtakegami@tmu.ac.jp (Daisuke Takegami)
E-mail: mizugu@tmu.ac.jp (Yoshikazu Mizuguchi)

Gen Nishijima
Research Center for Energy and Environmental Materials (GREEN), National Institute for Materials Science, Tsukuba, Japan

Shunsuke Kawakami, Tsutomu Nojima, Yuji Tsuchiya, Satoshi Awaji
Institute for Materials Research, Tohoku University, Sendai, Miyagi
E-mail: t.nojima@tohoku.ac.jp (Tsutomu Nojima)

Fuyuki Ando, Ken-ichi Uchida
Research Center for Magnetic and Spintronic Materials, National Institute for Materials Science, Tsukuba, Japan

Ken-ichi Uchida
Department of Advanced Materials Science, Graduate School of Frontier Sciences, The University of Tokyo, Kashiwa, Japan

Funding: TMU research fund for young scientists, JST-ERATO (No.: JPMJER2201), and JSPS-KAKENHI (No. 26K22467)

**Abstract text.**

Superconducting diodes (SDs) are devices that rectify supercurrent through the difference in the superconducting critical current ($I_c$) between forward and reverse current directions. For bringing such SDs closer to practical superconducting power electronics applications, a combination of both high rectification efficiency ($\eta$) and a large $I_c$ difference ($\Delta I_c$) is required. Here, we propose and demonstrate a coil-shaped superconducting wire device concept to induce a large SD effect, arising from the compensation between an external magnetic field ($H_{ext}$) and the field generated by the coil itself. From transport $I_c$ measurements on coil-shaped SDs based on a Sn10-Pb90 solder wire, we achieved a $\eta$ of 79% with a $\Delta I_c$ of 140 A at $H_{ext}$ = 500 Oe (coil #2) and a $\eta$ of 84% with a $\Delta I_c$ of 125 A at $H_{ext}$ = 530 Oe (coil #3), confirming high rectification values at large supercurrents. The large-current high-efficiency SDE in the coil-shaped SDs will be highly useful for developing the potential field of supercurrent power electronics.

## 1. Introduction

Superconducting diodes (SDs) exhibit nonreciprocity in their superconducting critical current ($I_c$); the $I_c$ differs between forward and reverse-direction current ($I_c^+ \neq |I_c^-|$).[1–4] Here, each direction is typically defined as $I_c^+ > |I_c^-|$. The SD effect (SDE) has attracted much attention since its observation in Nb-V-Ta superlattice in 2020, where it originates from the simultaneous breaking of spatial-inversion and time-reversal symmetries.[5] Since then, the SDE has been studied in a wide variety of systems, including unconventional-superconducting materials,[6–8] Josephson junctions,[9-11] and thin films and crystal flakes.[12,13] The most important parameters of SDs are the rectification efficiency ($\eta$), which is defined as $\eta = (I_c^+ - |I_c^-|) / (I_c^+ + |I_c^-|)$ and the critical-current difference ($\Delta I_c = I_c^+ - |I_c^-|$).[14] To enable practical applications in e.g., large-scale superconducting power electronics, the simultaneous achievement of a high $\eta$ (ideally perfect rectification of $\eta$ = 100%), and a large $\Delta I_c$ is essential. In low-current devices ($\Delta I_c$ < 10 mA), relatively high $\eta$ have been reported under standard measurement conditions with a DC current: $\eta$ = 65% in a Pt/V/EuS trilayer film and $\eta$ = 60% in a cuprate Josephson junction.[1,12,15]. A recent work also demonstrated perfect rectification in cuprate Josephson junctions with $\Delta I_c$ ~1 mA albeit using more complex experimental conditions including microwave irradiation.[16] $\eta$ > 50% was observed in a YBCO/PrBCO multilayer film but the observed $\Delta I_c$ was lower than 1 A.[17] However, when it comes to large supercurrents, so far only low $\eta$ have been obtained, with a device composed of a cuprate coated conductor and a pair of permanent magnets with antiparallel magnetizations obtaining $\eta$ ~ 15% with $\Delta I_c$ of 181 A [18]. Thus, to the best of our knowledge, no SD has yet simultaneously achieved a high $\eta$ (> 60%) and a large $\Delta I_c$ exceeding 100 A.

One of the potential applications for the above mentioned large-current SDs is high-current DC excitation, essential for superconducting magnets employed in emerging applications such as superconducting rotating machines and magnetic refrigeration, where classical current rectification introduces substantial thermal loads into the cryogenic environment.[19–21] Contactless flux-pump technologies have therefore been actively investigated to energize superconducting field coils.[18] Furthermore, if high-efficiency large current SDs working at $T$ = 77 K can be developed using cuprates, such devices could become essential for a future superconducting power grid system.[22] In this present work, to achieve both requirements of a high $\eta$ and a large $\Delta I_c$, we propose a coil-shaped superconducting wire device concept to induce a large SDE. We experimentally demonstrate that the device achieves performances of $\eta$ = 79% with a $\Delta I_c$ of 140 A and a $\eta$ = 84% with a $\Delta I_c$ of 125 A at $T$ = 4.2 K.

## 2. Results

### 2-1. Concept and Base Design

As illustrated in **Figure 1a**, a supercurrent flowing in a coil-shaped superconducting wire generates a magnetic field $\mathbf{H}_{in}(I)$ with a magnitude of $H_{in}(I)$ that is, according to the solenoid equation based on Ampère's circuital law, of $H_{in}(I) \sim IN/L$ with $I$ being the current, $N$ the number of turns, and $L$ the length of the coil or solenoid. Given an external magnetic field $\mathbf{H}_{ext}$ with a magnitude of $H_{ext}$ applied parallel to $\mathbf{H}_{in}$, the magnitude of total self-induced magnetic field inside the coil is expressed by $H_{total} \sim ||\mathbf{H}_{ext} \pm \mathbf{H}_{in}(I)||$, with the sign depending on the direction of the current. The critical current $I_c$ (at which $H_{total}$ is the critical field $H_c$) will then also depend on the direction of the current, inducing the SDE. As such, the present design utilizes the limitation of superconducting coils due to field generated by themselves as shown by Onnes[23] but aims to create the non-reciprocity by adding an external field. According to the working principle, if $H_{ext}$ is set close to $H_c$, a small current on the direction where the generated $\mathbf{H}_{in}$ is the same as $\mathbf{H}_{ext}$ ($I^-$) would immediately result in surpassing $H_c$, and thus, going to the normal-conducting phase, while a large current window on the opposite direction ($I^+$) would be available at the superconducting phase, theoretically providing very high rectification efficiencies close to $\eta \sim 100\%$. As summarized in Supporting Information, the SDE parameters within this concept can ideally be described as the following: $\Delta I_c = 2I_{c0}(H_{ext}/H_c)$ and $\eta = H_{ext}/H_c$ where $I_{c0}$ is the $I_c$ at $H_{ext} = 0$ Oe. Because of the simple working principle, the design strategy proposed here is highly scalable and versatile, allowing for bulk-scale SDs useful for large-scale superconductor applications.

To demonstrate this concept, we created a coil-shaped wire made of Sn-Pb solder (coil diode #1) and performed a four-terminal voltage ($V$)-$I$ measurement under $\mathbf{H}_{ext}$ (**Figure 1b**). $H_c$ of the Sn10-Pb90 solder is 570 Oe, close to that of pure Pb (See **Figure S1** for the magnetic characterization for a sample consisting of a small cylinder cut off from a similar wire) because the Sn-Pb solder is a phase-separated superconductor composed of Sn and Pb.[24–26] We note that, for a type-I superconductor, the demagnetization effect is crucial for discussing $H_c$,[27] and the effect can largely affect the diamagnetic property of coil-shaped samples. Since the precise evaluation of the demagnetization effect of complex or large samples such as the present device is complicated, we use the value of $H_c$ = 570 Oe estimated from a cylindrical sample with demagnetization-effect correction.

**Figure 1c** shows a $V$-$I$ measurement result at $H_{ext}$ = 250 Oe. Both curves display superconducting behavior below $I_c$, and the voltage dramatically shoots up due to the phase transition from superconducting to normal-conducting state. As expected from the working principle discussed above, for the reverse direction, a smaller $I_c^- = 47.42$ A is obtained, while on the forward direction it is of $I_c^+ = 122.70$ A, resulting in a $\Delta I_c$ of ~75 A and $\eta = 44\%$, experimentally proving the concept. As shown in **Figure S2**, at $H_{ext} > 250$ Oe, $I^+$ data could not be obtained due to the quenching at the wire outside the coil, detected by the additional voltage terminal (Figure 1a); the $I_c$ of the lead-wire part is consistent with our previous experiments on the same composition (see **Figure S3**).[26] Noticeably, the voltage near the coil was still zero with $I^+$ at $H_{ext} > 250$ Oe, which suggests that the $I_c^+$ of the coil part should have still increased with increasing $H_{ext}$, but the external quenching prevented it from reaching further. In **Figure 1d**, the experimentally measured $H_{ext}$ dependence of $I_c$ is shown. The results are well consistent and in good qualitative agreement with the predicted trends by the coil-shaped SDE parameter $\Delta I_c = 2I_{c0}(H_{ext}/H_c)$: for $I^-$, we observe a linearly decreasing trend converging to 0 at around $H_{ext} = H_c = 570$ Oe. As for $I_c^+$, it displays a linearly increasing trend, with a slope similar but opposite to that in $I^-$ given the additional current needed to first compensate for the opposing $H_{ext}$ and then, in addition, needed to reach $H_c$. Finally, the experimentally obtained $\Delta I_c$ and $\eta$ are shown in **Figure 1e** as a function of $H_{ext}$, with the largest values within the measured range (with $I^+$ lower than $I_c$ of the wire part) being $\Delta I_c = 75$ A and $\eta = 44\%$ at $H_{ext} = 250$ Oe.

In order to better understand the relation between the experimentally obtained $I_c$ and $H_c$, finite element method (FEM) simulations of electromagnetic fields were performed with the commercial software COMSOL Multiphysics® v. 6.4,[28] using the classical Maxwell equations built in the magnetic fields interface of the AC/DC module. The simulation results, summarized in **Figure S4** and **S5**, show that the coil $I_c$ is governed by local maximum field in the inner edge. As shown in Figure S4e, the experimental results shown in Figure 1 are nicely explained by total field at the wire inner edge. The simulation results confirm that the relevant parameters in coil-shaped SDs can be understood, on a first approximation, with simple electromagnetism for the Sn10-Pb90 wire. Although $I_c^+$ of coil #1 at $H_{ext} > 250$ Oe was not experimentally reached due to the quenching, a larger $\Delta I_c$ and a higher $\eta$ are expected at $H_{ext} > 250$ Oe as shown in Figure S5. Therefore, we next investigate the intrinsic SDE properties of an improved Sn10-Pb90 coil diode design for measurements at $H_{ext} > 250$ Oe by modifying the lead part to prevent quenching outside the coil.

## 2-2. High-Efficiency SDE with Superconducting Coil

In **Figure 2a**, we summarize the coil sample designed to prevent the lead-part quenching at $H_{ext} > 250$ Oe. We use pure-Cu lead (1.5 mm in diameter) having electrical resistance low enough for $I_c$ measurements at $I < 200$ A at $T = 4.2$ K. We measured two different coil diodes (#2 and #3) with opposite winding directions as shown in **Figure 2b** (coil diode #2 is right-handed, and #3 is left-handed). Both coil diodes are composed of the Sn10-Pb90 solder wire part (8.5 turns) and the outer pure-Cu wire parts. To avoid weak coupling between the Sn10-Pb90 and the Cu wires, the first and second turns on the outside of the coil are covered by soldering. As reported in Ref. 29, the normal-state resistivity ($\rho$) of Sn-Pb solder is $\rho \sim 3$ nΩm at $T = 4.2$ K, significantly higher than the typical value of pure Cu with $\rho \sim 0.01$ nΩm at $T = 4.2$ K. Therefore, the heating and/or the increase in resistance at the Cu part can be assumed to be negligible in the $I_c$ measurement with the examined current range. The voltage terminals are attached to the Sn10-Pb90 wire at about 7.4 cm apart.

**Figure 2c and 2d** show the selected current dependence of voltage ($V$-$I$) at $T = 4.2$ K for coil diodes #2 and #3. The other $V$-$I$ data are displayed in **Figure S6** and **S7**. We determine $I_c$ with the 1μV/cm criterion. For coil #2, the highest $\eta$ of 79% is observed with $\Delta I_c = 140$ A at $H_{ext} = 500$ Oe. For coil #3, the highest $\eta$ of 84% is observed with $\Delta I_c = 125$ A at $H_{ext} = 530$ Oe. Although the obtained SDE properties are slightly different between #2 and #3, the results at all the examined fields are comparable. This suggests that the winding direction, right-handed or left-handed, does not affect the working performance of the coil-shaped SDs. In **Figure 3**, $I_c^+$ and $I_c^-$ for coils #2 (Figure 3a) and #3 (Figure 3b) are plotted as a function of $H_{ext}$. Owing to the improved device design with the pure-Cu lead parts, which avoids quenching outside the coil, $I_c^+$ at higher $H_{ext}$ are reachable in these coils. For both coils, $|I_c^-|$ linearly decreases with increasing $H_{ext}$. $I_c^+$ shows a monotonous increase with increasing $H_{ext}$ up to 400 Oe, but the increase is moderated at $H_{ext} > 400$ Oe, showing a deviation from the idealized sample and measurement set-up, shown with the dashed lines, suggesting the presence of additional constraints to reach the ideal rectification. Nonetheless, experimentally, the largest $I_c^+$ is observed at $H_{ext} \sim 500$ Oe for both coils with $\eta = 79\%$ and 84%, both cases record-high values among the SDE studies with conventional measurement conditions with a DC current.

## 3. Discussion

To compare our data with other SDs' performance, $\eta$ (higher than 60% for our coil-shaped SDs) is plotted as a function of $\Delta I_c$ and working temperature in **Figure 4a and 4b**. In Figure 4a, we can clearly see that the present work achieved a significant improvement in $\eta$ for large $\Delta I_c$ devices. As for a broader comparison including lower $\Delta I_c$, we observe that only the microwave-irradiated designs, which require much more complex operating conditions, have reported higher $\eta$, with our device showing a significant improvement also with respect to other conventional SDs. Our coil-shaped SDs show a high $\eta$ with a large $\Delta I_c$ exceeding 100 A under low operating magnetic field of ~500 Oe that can be readily achieved using an ordinary permanent magnet, which are suitable for practical large-current superconductor applications. According to the working principle, $\Delta I_c$ becomes larger with increasing zero-field $I_c$, which can be enhanced by a reduction of turn number and/or an increase in $H_c$ (or irreversible field of type-II superconductors) by changing the coil material. However, the current working temperature is $T = 4.2$ K, and there is room for improvement for $\eta > 60\%$ and $T > 4.2$ K; working temperatures of $T = 20$ and 77 K are important for practical applications with liquid hydrogen and nitrogen. This suggests a potential direction for further exploration using the same concept of coil-shaped SDs using superconductors different from the Sn10-Pb90 solder such as $Nb_3Sn$ or cuprate wires that could result in higher working temperatures.

At the same time, the suppression of $I_c^+$ and deviation from the theoretical expectations for increased $H_{ext}$ shows that there might still be room for design optimization and improvement. In the present work, we do not have a clear answer to the $I_c^+$ deviation from the theoretical value near $H_c$ within the present experiments, but $I_c^+$ might be sensitive to small deviations in the field direction or the coil structure including wire diameter or shape. These are thus parameters that could further be explored in future designs. Further improvement of SDE properties including working temperature will improve the performance of various superconducting devices driven with a large current and be an essential tool for developing future superconducting power electronics.[19-22,30,31]

**Figures**

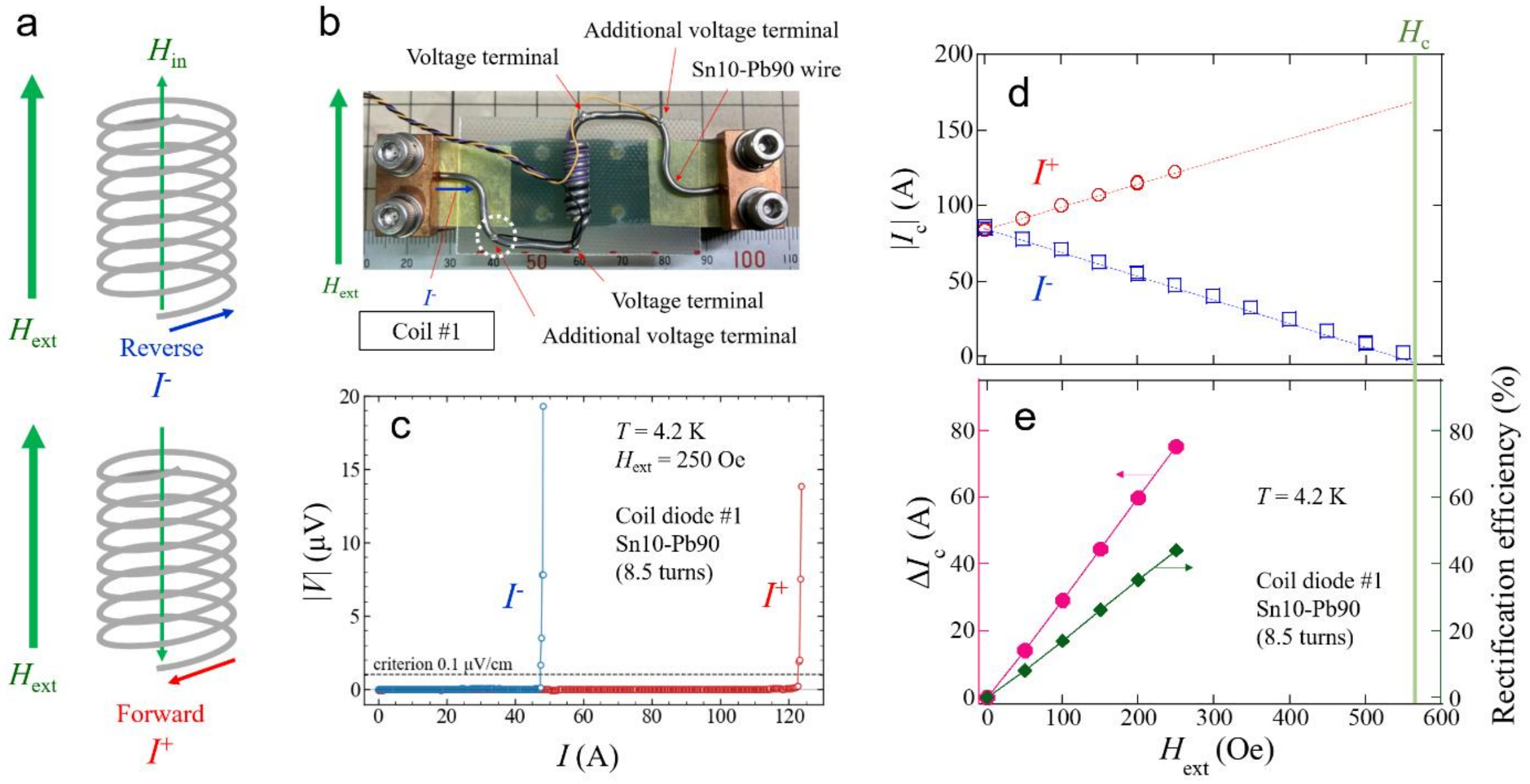

**Figure 1.** Concept and $I_c$ measurements for coil diode (#1). a) External field ($H_{ext}$) and inductive field ($H_{in}$) by the supercurrent with forward and reverse directions ($I^+$ and $I^-$). b) A photograph of the measurement configuration with Sn10-Pb90 coil sample (#1) and current and voltage terminals. Additional voltage terminals are prepared to investigate the problem of quenching the wire outside the coil (see dashed white circle). c) $V$-$I$ data taken with $I^+$ and $I^-$ at $T = 4.2$ K and $H_{ext} = 250$ Oe, plotted with red and blue circles, respectively. d) $H_{ext}$ dependence of $I_c$ at $T = 4.2$ K. The dashed lines are theoretical values of $I_c^+$ and $I_c^-$ estimated from $I_{c0}$ and $H_c = 570$ Oe (see Supporting Information for the formula). $H_{ext}$ dependence of expected $\Delta I_c$ and rectification efficiency.

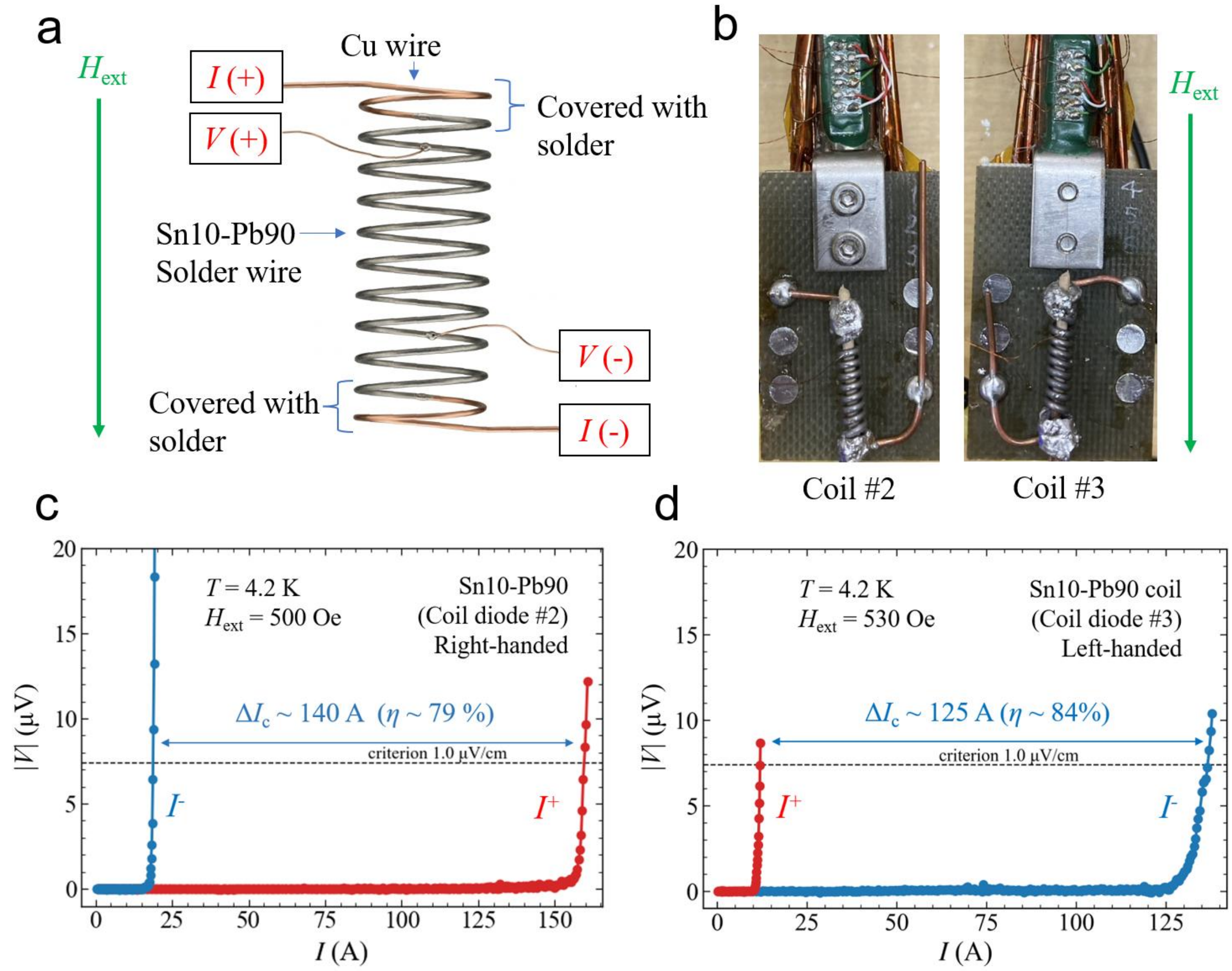

**Figure 2.** Critical current experiments. a) Schematic image of the concept. b) Sample photographs for coils with opposite winding directions. c,d) Typical data for the current dependence of voltage for the right- and left-handed coil diodes (#2 and #3). Here, the criterion for estimating $I_c$ is 1 μV/cm. $I^+$ and $I^-$ denote the current directions.

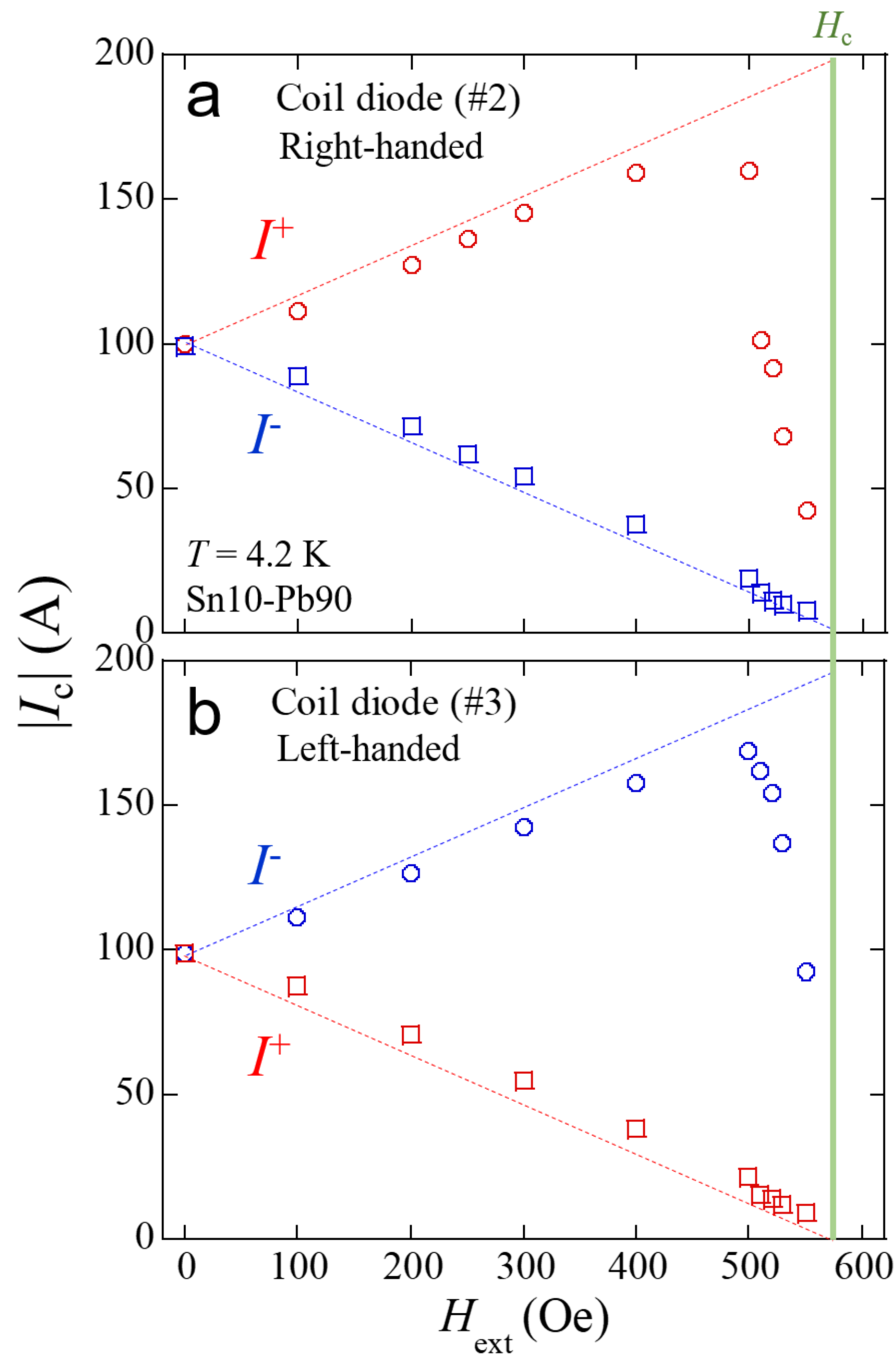

**Figure 3.** External field dependence of $I_c$ for the coil diodes, a) #2 and b) #3. The dashed lines are theoretical values estimated from $I_{c0}$ and $H_c$ = 570 Oe.

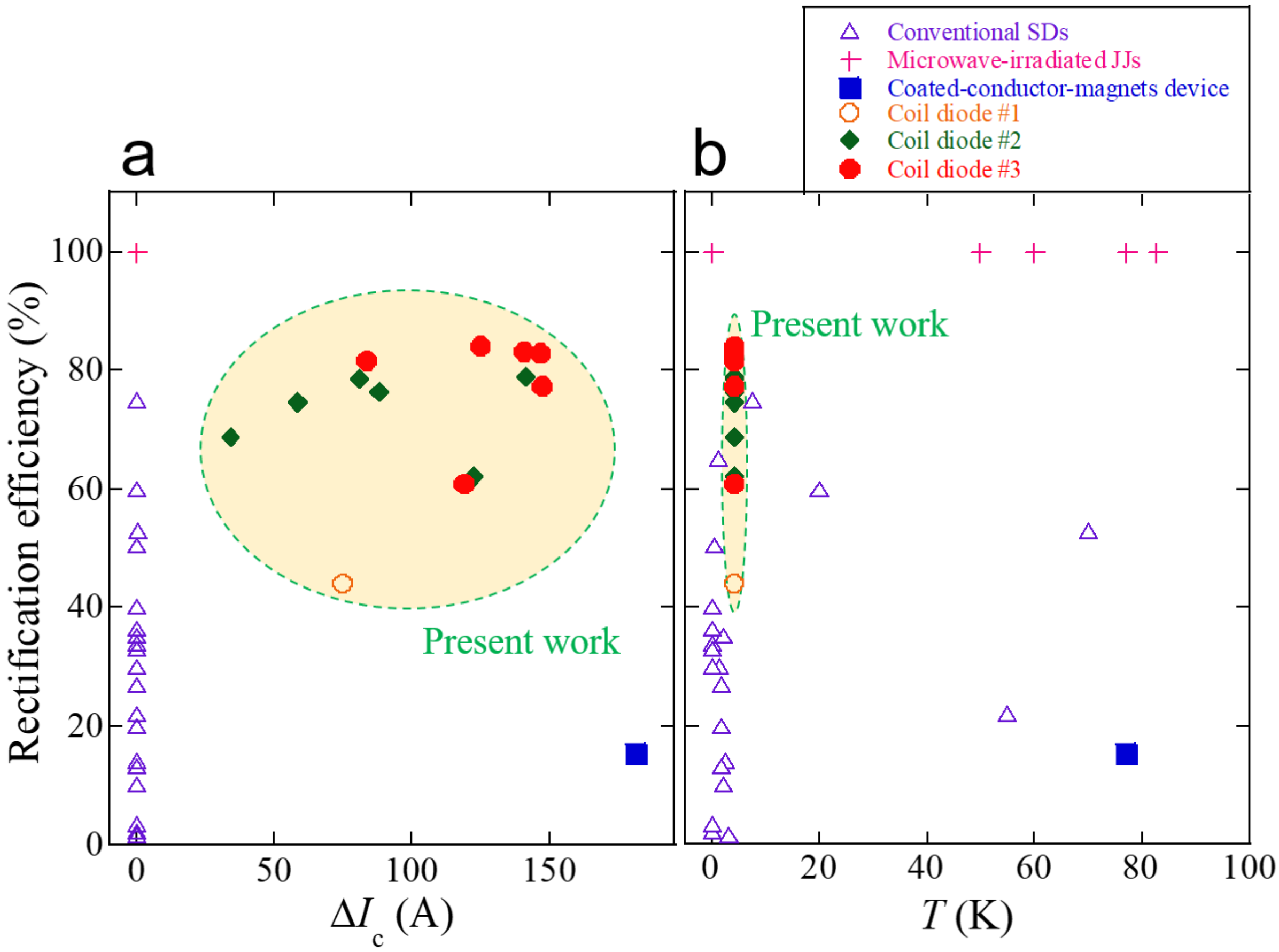

**Figure 4.** Comparison of the rectification efficiency as functions of a) $\Delta I_c$ and b) experiment temperature. The efficiency data for the conventional SDs are mainly provided from J. Ma et al.[1] We included a record-high data for SD without microwave irradiation reported in arXiv.[32]

## 4. Conclusion

In this study, we propose coil-shaped SD as a new design strategy to achieve a giant supercurrent rectification and experimentally demonstrate that the coil-shaped Sn10-Pb90 wire shows SDE. The development of SDs had been so far mainly limited to the nano- and micro-scale given their use of thin films and nano- and micro-fabrication approaches, resulting in a focus on small-scale potential applications such as quantum computing. The bulk-scale SDs presented in this work features a simpler design and higher efficiency, and expands the possibilities of bulk-scale SDs to large-scale applications. We demonstrate that a coil-shaped Sn10-Pb90 SD with pure-Cu lead parts exhibits high $\eta$ and large $\Delta I_c$. We experimentally observed a large enhancement of $I_c^+$ by the field cancellation principle at higher $H_{ext}$. The highest $\eta$ values are 79% with $\Delta I_c$ = 140 A (coil #2) and 84% with $\Delta I_c$ = 125

A (coil #3), marking a significant improvement to large supercurrent SDs as well as to overall conventional SDs. High-performance SDs with a large $\Delta I_c$ will open new pathway for versatile superconducting power applications.

## 5. Materials and Methods

5.1. Sample information

The measurements were performed using a Sn10-Pb90 (wt.%) solder cylindrical wire with a diameter of 1.65 mm (Sasaki Solder Industry). For coil #2 and #3, the solder wire was connected with pure-Cu wires (99.9%) with a diameter of 1.5 mm (Nilaco). The diameter of pure-Cu wire is enough for $I_c$ measurements with $I < 200$ A. The solder wire and the Cu wire were connected by soldering, and the overlapping parts were covered by solders to enhance the joint $I_c$.

5.2. $I_c$ measurement for coil diode #1

The transport $I_c$ for coil #1 was measured using the cryostat system in NIMS by a standard four-probe method using a Keithley 2182A nanovoltmeter (see Figure 1b for the measured sample). Continuous DC current was used for the measurement, and the sweep speed from zero to the highest current was 20-30 s. Both ends of the solder wire were mechanically fixed to the current terminals. The voltage taps were bonded using low-melting-point solder (U-Alloy 95) and protected with adhesive tape. The distance between the voltage taps near the coil ends was about 10.6 cm. The applied $H_{ext}$ was parallel to the coil-length direction. Measurements were performed in liquid He, and the $I_c$ was evaluated using a criterion of 0.1 $\mu$Vcm$^{-1}$.

5.3. $I_c$ measurement for coil diodes #2 and #3

The transport $I_c$ for coil diodes #2 and #3 was measured using the cryostat system in IMR, Tohoku University, by a standard four-probe method using a Keithley 2182A nanovoltmeter (see Figure 2b). Both ends of the solder coil were connected with the Cu wires by soldering using a Sn40-Pb60 solder. The sample was fixed to the current terminals by soldering using a low-melting-point solder (In52-Sn48). This In52-Sn48 solder was also used for attaching the voltage terminals to the Sn10-Pb90 coil part. The sweep rate of the DC current from zero to the maximum value was 30 sec. The applied field ($H_{ext}$) was almost parallel to the coil-length direction, and the actual field was measured by a Hall device. The measurements were performed in liquid He, and the $I_c$ was evaluated using a criterion of 1 $\mu Vcm^{-1}$.

**Acknowledgements**

T. Ichikawa and D. Takegami equally contributed to this work. The authors thank M. Mashiko, Y. Watanabe, and T. Ozaki for supports and discussion on this research. The authors thank J. Ma and X. Lin for providing the source data for the efficiency plots published in their review article. This work was partly supported by TMU research fund for young scientists, JST-ERATO (No.: JPMJER2201), and JSPS-KAKENHI (No. 26K22467).

**Data Availability Statement**

The data that support the findings of this study are available from the corresponding author upon reasonable request.

**Conflict of Interest**

The authors declare no conflict of interest.

# Supporting Information

**Expected SDE performances of coil superconducting diodes**

We demonstrate that the ideal critical current difference ($\Delta I_c$) and rectification efficiency ($\eta$) are determined by $I_{c0}$ ($I_c$ at $H_{ext}$ = 0 Oe) and the critical current $H_c$. $I_{c0}$ depends on the coil shape and the turn number. Because the generated field ($H_{coil}$) is proportional to current in a coil, $I_c^-$ is described as $I_c^- = I_{c0}[1 - (H_{ext}/H_c)]$. Similarly, $I_c^+$ is described as $I_c^+ = I_{c0}[1 + (H_{ext}/H_c)]$. Therefore, the ideal $\Delta I_c(H_{ext})$ and $\eta(H_{ext})$ are described as following.

$$\Delta I_c = 2I_{c0}\left(\frac{H_{ext}}{H_c}\right)$$

$$\eta = \left(\frac{H_{ext}}{H_c}\right)$$

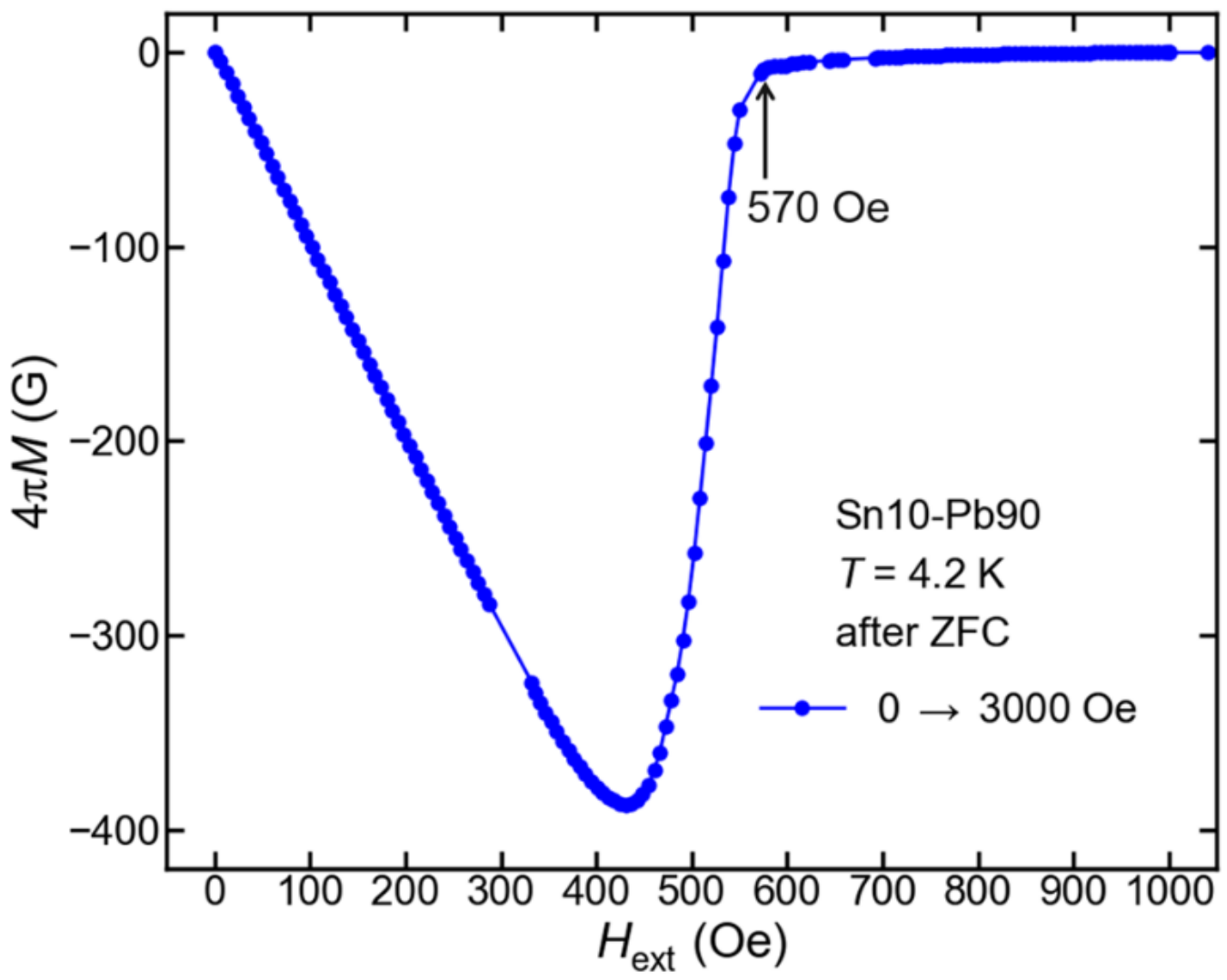

**Fig. S1.** $H_{ext}$ dependence of magnetization for Sn10-Pb90 solder at $T$ = 4.2 K. Original data has been published in T. Ichikawa et al., Jpn. J. Appl. Phys. 65 100904 (2026). Magnetization was measured using a superconducting quantum interference device (SQUID) magnetometer MPMS3 (Quantum Design) under various fields in the SQUID-VSM mode.

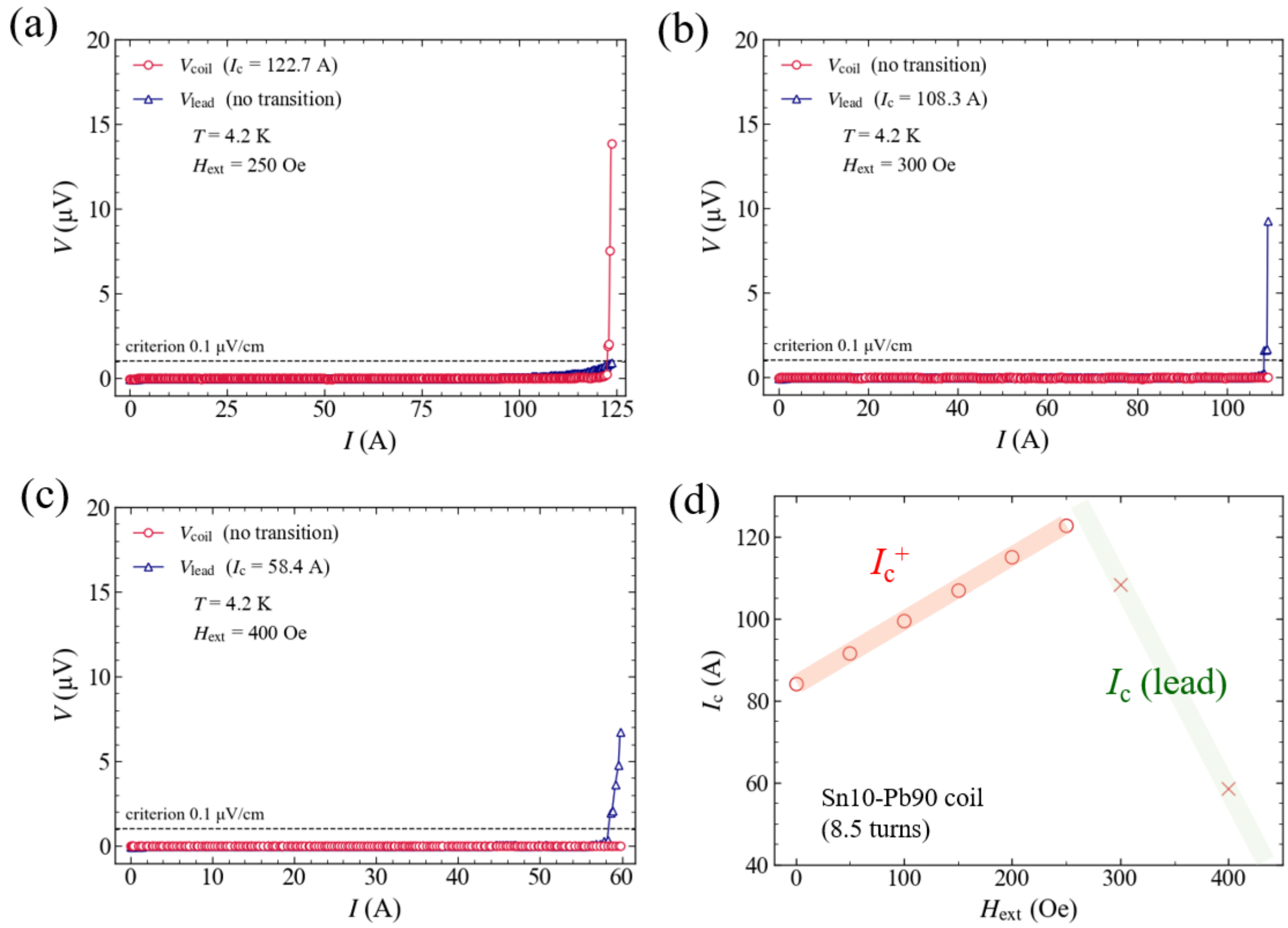

**Fig. S2.** (a-c) *V-I* measured near the coil and the lead part (see Figure 1a). The sudden increase in blue-triangle data indicates quench at the wire part. (d) $H_{ext}$ dependence of critical currents, $I_c^+$ and $I_c$ (lead), for Sn10-Pb90 solder.

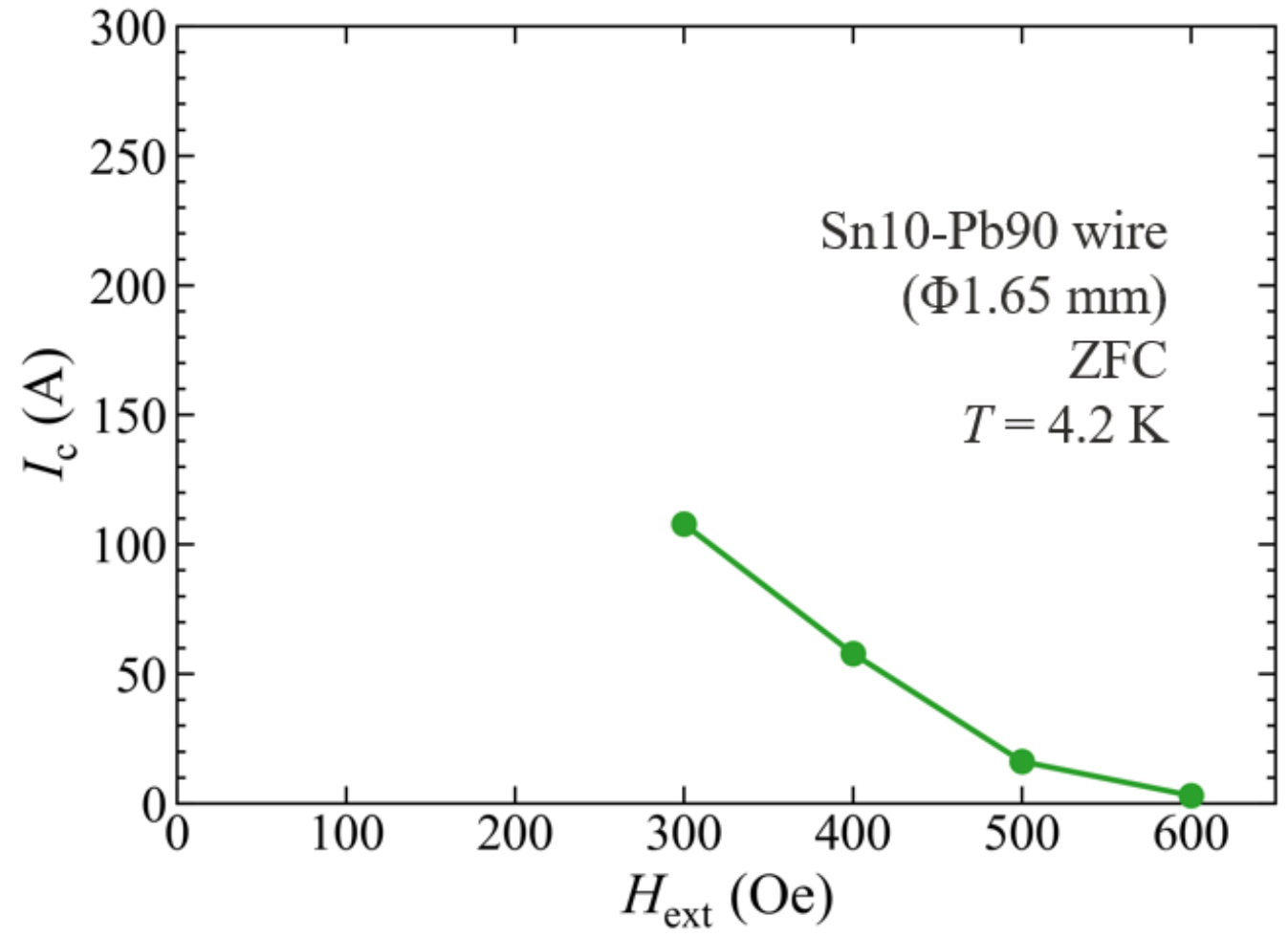

**Fig. S3.** Hext dependence of $I_c$ for a straight-shaped Sn10-Pb90 wire. The original data has been published in T. Ichikawa et al., Jpn. J. Appl. Phys. 65 100904 (2026).

**Finite element method simulations**

Finite element method (FEM) simulations of electromagnetic fields were performed with the commercial software COMSOL Multiphysics® v. 6.4, using the classical Maxwell equations built in the magnetic fields interface of the AC/DC module. The core part of the coil was modelled using a solenoid with an inner diameter of 2 mm, a wire diameter of 1.65 mm and an axial pitch of 2.5 mm, resulting in a total length of around 21 mm. Additional cylindrical and toroidal elements were attached to the ends to simulate the straight and curved sections of the contacts respectively, as shown in Figure S4a, although the details on the external elements did not result in significant differences in the area of interest within and surrounding the coil. A constant intensity of $I$ was set entering / exiting through the contacts as boundary conditions, and the current density within the wire was simulated based on Ohm's law. We note that the present calculations are a first approximation based on the classical, normal conduction equations. An electric permittivity of $\varepsilon_r = 10^{10}$, relative permeability of $\mu_r = 10^{-5}$and resistivity of $\rho = 10^{-15}\Omega$m were used for the wire. A box of empty space of 4 cm × 4 cm × 20 cm was defined centered at the coil to allow for enough space for the out magnetic flux lines to propagate. The external field was simulated by setting homogeneous magnetic field along the z direction on the top and bottom surfaces of the box. The remaining faces, and in the case of zero field calculations, all faces, were set as magnetic insulators, i.e. with the perpendicular component of the field to be 0.

Figure S4b–S4d display the spatial distribution of the magnetic field using the parameters $H_{ext}$ = 250 Oe and $I^{\pm}$ = ±80 A. For $I^{-}$, the magnetic field generated inside the coil due to the current is on the same direction as the external field, with their magnitudes adding up, and resulting in a field well above the experimentally determined $H_c$, as observed in the magnetic field along the central axis of the coil, represented in Fig. 3b. Meanwhile, for $I^{+}$, given the opposite direction of the magnetic field generated by coil and $H_{ext}$, the field inside is still well below $H_c$, of the order of $|H_{coil} - H_{ext}|$ where $H_{coil}$ denotes $H$ generated by the coil. This would result in the wire remaining superconducting only for $I^{+}$, as conceptualized and experimentally demonstrated above. We observe that the field, which outside the coil is of $H_{ext}$, decreases near the boundary of the coil, dropping to 0 and then increasing again in magnitude after $H_{coil}$ surpasses $H_{ext}$. Additional non-trivial asymmetric behavior is present between $I^{\pm}$, due to the distortion of the external field in presence of the coil. Figure S4d displays a colourmap of the magnetic field magnitude distribution generated by the model coil on top of an external field. Outside the coil, we observe uniformly low values due to the magnetic field lines being highly dispersed through all the available space. On the other hand, the values inside the coil are much higher due to the concentration of the field. The field inside is far from homogeneous, with the field in the vicinity of the wires being significantly higher. In the cross-sectional cut, shown in Figure S4c, the field at the inner edges of the coil reaches values around 1.5 times higher than that in the center. This behavior is to be expected, given the $1/d$ ($d$ being the distance) dependence of the magnetic field generated by a straight wire section. With this behavior in mind, we can now calculate for a range of values in order to obtain the theoretically expected $H_{ext}$ dependence of $I_c$ according to our finite element model. In Figure S4e, we show a color map of the $H_{total}$ at the inner edge of the wires forming the coil, with the dashed region denoting the parameter space where $H_{total} > H_c$ =570 Oe. The cutoff boundary where $H_{total} = H_c$ (corresponding to $I_c$), shown by the green line, is in excellent agreement with the experimentally measured values, showing that the magnetic field magnitude at the inner edge provides a much better prediction of the characteristics of coil-shaped SD devices. The agreement of the $I_c$ values obtained with a simplified FEM analysis shows that the working parameters of the coil-shaped SD can be estimated with low-cost computational models using classical electromagnetism. Meanwhile, the cutoff values and $I_c$ obtained from the center of the coil in our model (dotted purple line in Fig. S4e) are in full

agreement with the results obtained with the solenoid equation, indicating, thus, consistency of our FEM analysis with the theoretical equations.

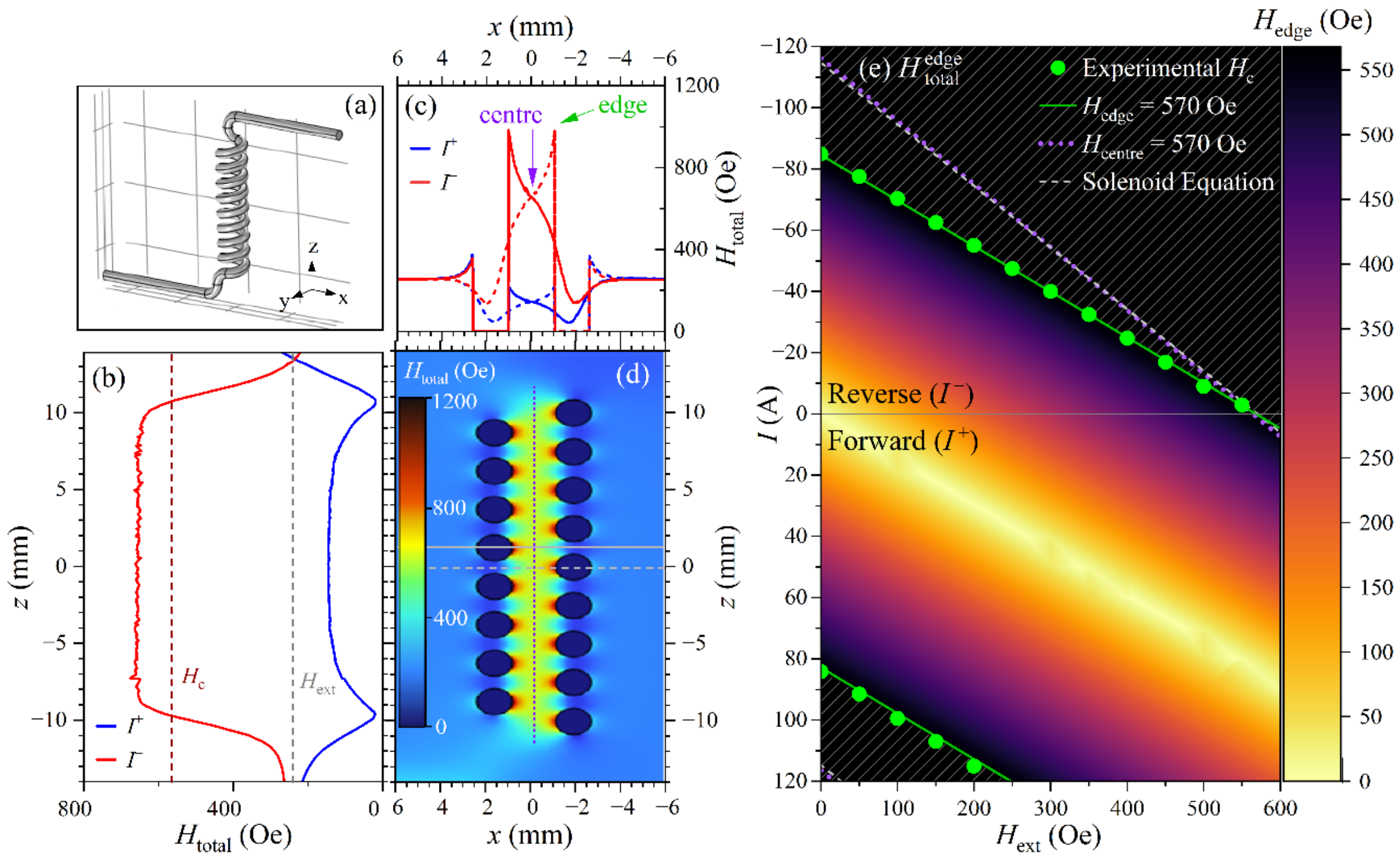

**Fig. S4.** Finite element simulations for the spatial distribution of the magnetic field magnitude around a coil-shaped wire. (a) 3D rendering of the model coil used. (b,c,d) Calculated $H_{total}$ along (b) the center axis of the coil and (c) across cross-sectional cuts of the model, as displayed in (d) the 2D projection of the model on the *xz*-plane cutting the middle of the coil. (e) Color map depicting the $H_{edge}$ as a function of $H_{ext}$ and $I$, together with the $H_{ext}$ dependence of $I_c$ at the inner edge (green solid line) and center of the coil (purple dotted line). The experimental points from Figure 1c (green circles) are shown for comparison.

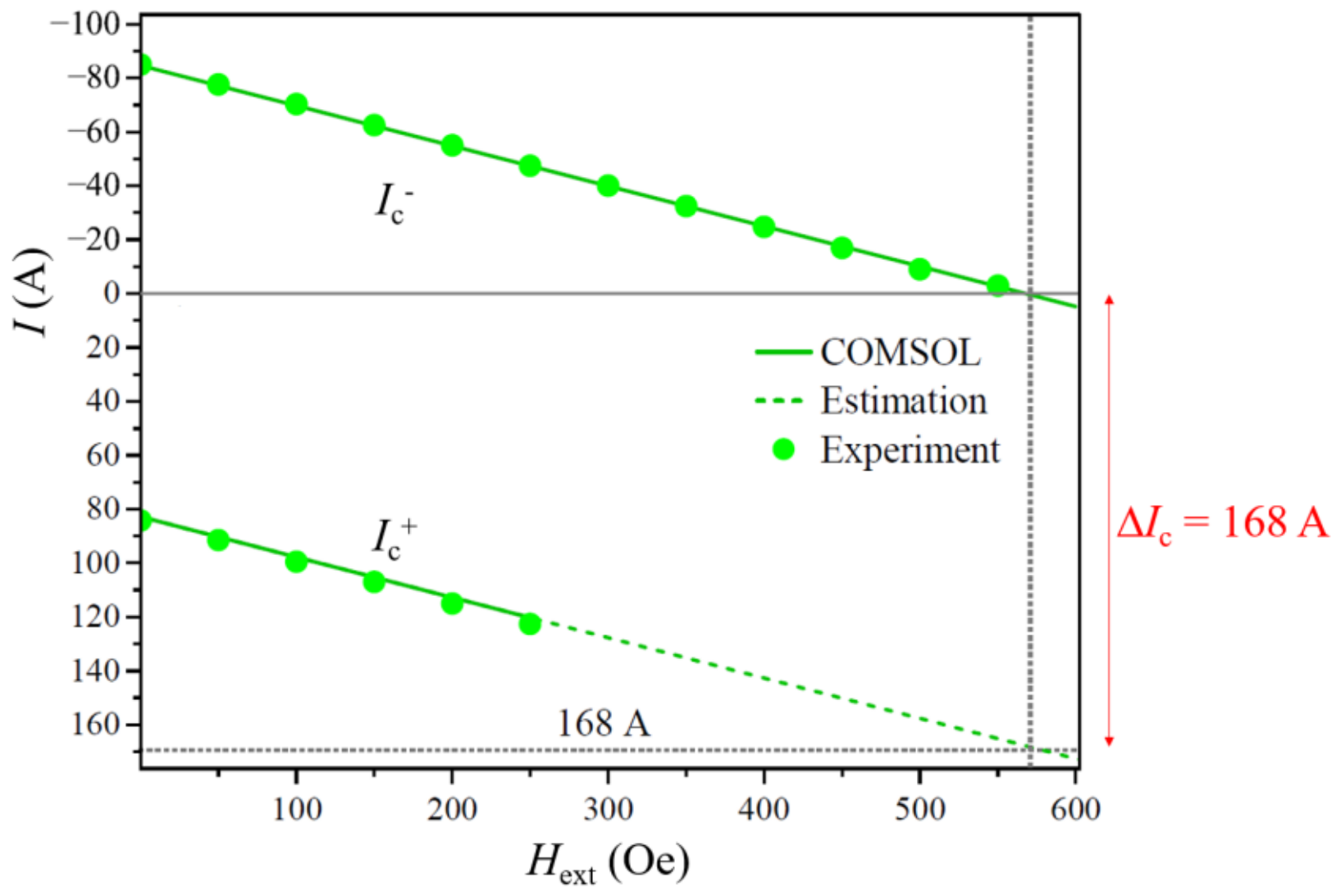

**Fig. S5.** $H_{ext}$ dependence of critical currents, $I_c^+$ and $I_c^-$, for Sn10-Pb90 solder with $H_c$ = 570 Oe at $T$ = 4.2 K.

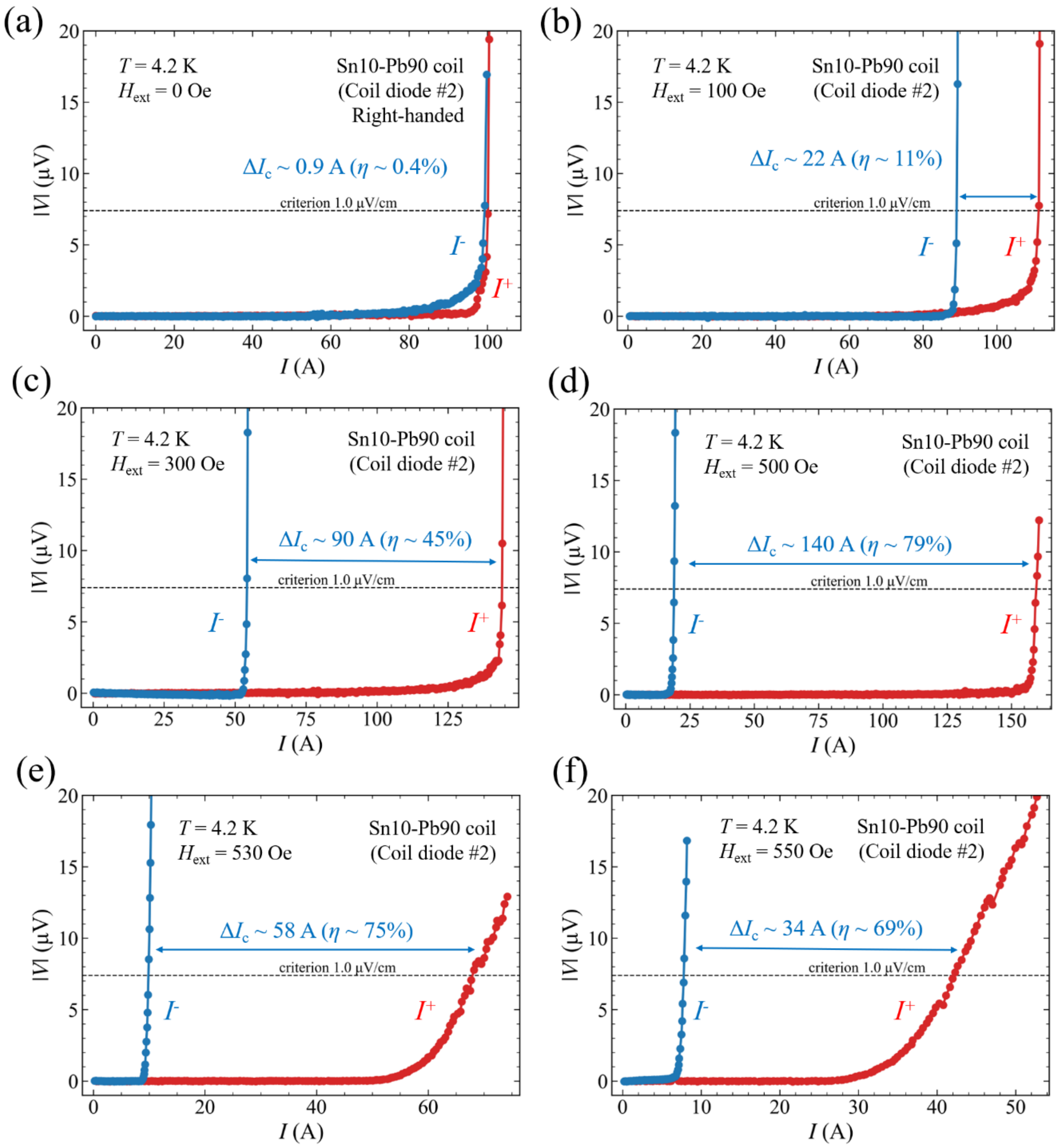

**Fig. S6.** $V$-$I$ data taken at various $H_{\mathrm{ext}}$ for coil diode #2.

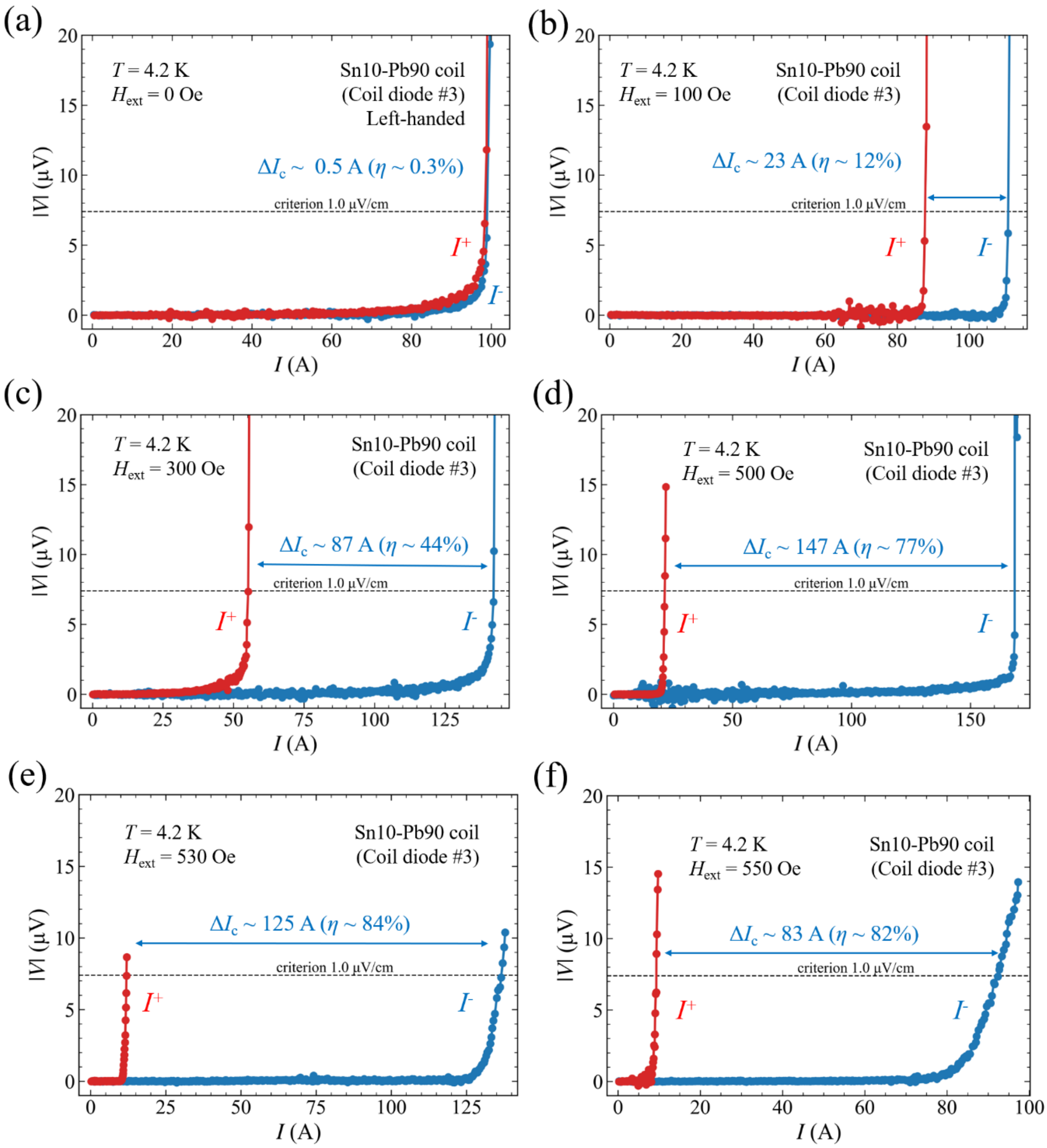

**Fig. S7.** *V*-*I* data taken at various $H_{ext}$ for coil diode #3.